\pdfoutput=1
\documentclass[12pt]{iopart}

\usepackage{soul}
\usepackage{graphicx}
\usepackage{latexsym}
\expandafter\let\csname equation*\endcsname\relax
\usepackage{xcolor}
\expandafter\let\csname endequation*\endcsname\relax
\usepackage{amssymb}
\usepackage{amsmath}
\usepackage{mathrsfs} 
\usepackage{todonotes}
\usepackage{caption}
\usepackage{subcaption}
\usepackage{orcidlink}
\usepackage{booktabs}
\usepackage{float}
\usepackage{xcolor}

\begin{document}

\title{Observational Constraints on Emergent Fractional Fractal Cosmology}

\author{R. Jalalzadeh$^{1}$\footnote{Author to whom any correspondence should be addressed} \orcidlink{0000-0002-6110-3981},  H. Moradpour$^1$ \orcidlink{0000-0003-0941-8422}, and S. Jalalzadeh$^{2,3,4}$ \orcidlink{0000-0003-4854-2960}}
\address{$^1$ Research Institute for Astronomy and Astrophysics of Maragha (RIAAM),
 P. O. Box 55136-553, Maragheh, Iran}
\address{$^2$ Izmir Institute of Technology, Department of Physics, Urla, 35430, Izmir, Türkiye}
\address{$^3$ Center for Theoretical Physics, Khazar University, 41 Mahsati Str., Baku, AZ1096, Azerbaijan}
\address{$^4$ Department of Physics, Dogus University, Dudullu-Ümraniye, 34775 Istanbul, Türkiye}
 
 \ead{r.jalalzadeh@riaam.ac.ir, h.moradpour@riaam.ac.ir, shahramjalalzadeh@iyte.edu.tr}

 \vspace{10pt}
\begin{indented}
\item[]
\end{indented}

\begin{abstract}
	We constrain the Emergent Fractional Fractal (EFF) cosmological model through a joint likelihood analysis of recent cosmological observations at the background and perturbation levels. In this framework, an effective fractal dimension $d$ is introduced to parameterize possible fractional deviations from the standard cosmological model. We consider three combinations of datasets: (i) late-time (LT) observations including PantheonPlus Type Ia supernovae, $H(z)$ measurements, and growth-rate measurements $f\sigma_8$; (ii) LT combined with DESI DR2 BAO and Big Bang nucleosynthesis (BBN); and (iii) LT combined with DESI DR2 BAO and CMB distance priors. With the inclusion of CMB distance priors, the fractal dimension is constrained to $d = 2.0004^{+0.0006}_{-0.0003}$ at the $1\sigma$ confidence level. Model comparison using the Akaike Information Criterion (AIC) shows that the EFF and $\Lambda$CDM models fit the observational data equally well, while the Bayesian Information Criterion (BIC) favors the simpler $\Lambda$CDM model because of its smaller parameter space. These results show that current cosmological observations place strong constraints on fractal extensions of the standard cosmological framework and possible deviations from the $\Lambda$CDM model.
\end{abstract}

\section{Introduction}\label{Introduction}

It is important to consider both the ultraviolet and infrared regimes when studying quantum gravity. In this context, gravity and quantum mechanics are combined to describe the Universe across all scales, from the Planck scale to cosmological distances. Asymptotic safety~\cite{Reuter:1998}, loop quantum gravity~\cite{Rovelli:2008}, and string theory~\cite{Zwiebach:2004} are among the main approaches in this area. These approaches aim to connect general relativity with quantum mechanics and to address key issues such as singularities~\cite{Jalalzadeh:2025}, renormalizability~\cite{Calcagni:2010}, and the black hole information paradox~\cite{Hawking:1976}. In addition to these efforts, other innovative approaches have gained attention in recent years, particularly frameworks that use fractional calculus and fractal spacetimes to model nonlocal and memory-dependent quantum effects~\cite{Moniz:2020emn, Jalalzadeh:2022, Chen:2024}. These theories provide effective tools for investigating modifications of standard cosmology in the ultraviolet and infrared regimes~\cite{JalalzadehMoniz:2022, JalalzadehCosta:2022}.

Fractional fractal cosmological models provide a framework for describing spacetime with an effective nonsmooth geometry at the Planck scale~\cite{Calcagni:2017, Costa:2023, Jalalzadeh:2024}. In such approaches, spacetime may effectively exhibit fractal like properties across different physical scales. The use of fractional calculus in cosmology leads to an effective spacetime with a fractional, rather than integer dimension. Fractional derivatives, such as the Riesz derivative, allow the incorporation of long range effects and memory effects in cosmic evolution. This framework is used to study phenomena such as power-law inflation, the evolution of cosmological parameters in the early Universe, and the distribution of dark matter~\cite{ElNabulsi:2013, Rasouli:2022, Benetti:2024}. Fractional frameworks also provide methods for estimating the cosmological constant and for investigating modified Newtonian dynamics (MOND) at galactic scales, as a consequence of the fractional structure of spacetime~\cite{Landim:2021, Giusti:2020}.

The emergent fractional fractal (EFF) cosmological model is formulated by introducing fractional derivatives into the Wheeler--DeWitt equation~\cite{daSilvaJunior:2023}. In this framework, the entropy associated with a fractal horizon controls the evolution of the Universe, and the standard cosmic expansion is modified by the presence of a single additional degree of freedom, namely the effective fractal dimension $d$. In the limit $d=2$, the model reduces to the standard $\Lambda$CDM cosmology.

The EFF framework aligns with Padmanabhan’s concept of emergent cosmic space~\cite{Padmanabhan:2012}. In this approach, the expansion of the Universe is interpreted as the continuous emergence of space over cosmic time. Cosmic expansion is driven by the difference between the surface and bulk degrees of freedom, $N_{\text{sur}} - N_{\text{bulk}}$.
Within this holographic paradigm, thermodynamic quantities such as horizon entropy and temperature are closely related to gravitational and bulk degrees of freedom and play a central role in cosmic evolution. By combining Padmanabhan’s emergent space concept with the continuity equation for a perfect fluid, the Friedmann and Raychaudhuri equations can be derived, as demonstrated in previous studies~\cite{Cai:2012, Tu:2013, Hashemi:2015, Yuan:2013, Moradpour:2016, Chen:2022}.

Current astronomical observations allow cosmological models to be tested over a wide range of redshifts, from the present epoch to the earliest stages of the Universe, thereby providing stringent tests of their validity. The original formulation of the EFF model suggested that the modified cosmological dynamics induced by the fractal structure of spacetime could mimic some effects usually attributed to dark matter. However, such a possibility must be examined carefully using observational data. Precision cosmological measurements, particularly those related to the expansion history of the Universe and the growth of large-scale structure, provide strong constraints on any deviation from the standard $\Lambda$CDM paradigm and allow us to test whether the EFF framework can reproduce the observed cosmological dynamics.

The EFF cosmology can be interpreted as an extension of the standard $\Lambda$CDM model, where the parameter $d$ characterizes the effective fractal structure of spacetime and quantifies deviations from the standard Friedmann dynamics. The model reduces exactly to $\Lambda$CDM in the limit $d=2$, ensuring consistency with the conventional cosmological framework. For $d\neq2$, the modified continuity equation changes the scaling of matter and radiation densities, which affects both the expansion history and the growth of cosmic structures. Consequently, late-time observations such as Type Ia supernovae, $H(z)$ measurements, and growth-rate measurements $f\sigma_8$ constrain the low-redshift behavior of the model, while allowing mild degeneracies between $d$ and parameters such as $H_0$, $\Omega_m$, and $\sigma_8$. These degeneracies are significantly reduced when geometric probes such as BAO and CMB distance priors are included, since they constrain the background expansion and the sound horizon scale, leading to tighter limits on deviations from $\Lambda$CDM.

In this work, we constrain the EFF model previously developed in~\cite{daSilvaJunior:2023} using a combination of cosmological datasets, including baryon acoustic oscillations (BAO)~\cite{DESI:2025}, PantheonPlus Type Ia supernovae~\cite{Scolnic:2022}, $H(z)$ measurements from cosmic chronometers and radial BAO analyses~\cite{Tamri:2026, Mohebi:2026}, Planck CMB distance priors~\cite{chen:2019}, Big Bang nucleosynthesis~\cite{pitrou2018precision}, and growth-rate measurements~\cite{Mohebi:2026}.
To investigate the impact of different observational probes on the model parameters, we consider three dataset combinations: late-time (LT) probes including SN, $H(z)$, and $f\sigma_8$; LT + DESI DR2 BAO + BBN; and LT + DESI DR2 BAO + CMB.
We aim to examine the observational viability of the EFF framework and assess whether it can describe current cosmological observations as successfully as the standard $\Lambda$CDM model.

This paper is organized as follows: In Section~\ref{theory}, we theoretically investigate the EFF model and discuss the cosmology derived from fractional fractal entropy. The statistical methods and numerical results are presented in Sections~\ref{obser} and~\ref{num}, respectively. Finally, in Section~\ref{con}, we conclude our main results.

\section{The theory of  Emergent Fractional Fractal Cosmic Space}\label{theory}

In this section, we provide a brief overview of the EFF model originally proposed in~\cite{daSilvaJunior:2023}. This framework is based on fractional quantum gravity, in which the event horizon of a black hole can be described as a random fractal surface with a fractal dimension between 2 and 3. Extending this idea to cosmology, the apparent horizon of the Universe is assumed to exhibit a similar fractal structure.

Following Padmanabhan’s emergent space approach, the expansion of the Universe is described as the result of a difference between the degrees of freedom on the horizon’s surface and those inside the bulk. 
\begin{equation}
	\frac{dV}{dt}=L_\text{P}^2(N_\text{sur}-N_\text{bulk}),
\end{equation}
When the horizon has a fractal structure, this balance changes, leading to modified gravitational equations. Therefore, the usual Friedmann and Raychaudhuri equations include a correction that depends on the effective fractal dimension, which changes how the energy densities evolve.

The fractional origin of the model can be traced back to the WDW equation.
The fractional counterpart of the WDW equation in a $v$-dimensional configuration space with coordinates $q^\alpha$ ($\alpha=0,\dots,v$) is given by
\begin{equation}\label{WDW_frac}
	\left\{\frac{M_\text{P}^{2-\alpha}}{2}(-\Box)^{\frac{\alpha}{2}}-U(q^\nu)\right\}\Psi(q^\nu)=0,
\end{equation}
where $(-\Box)^{\alpha/2}$,  $\alpha\in(1,2)$, and $M_\text{P}$ are the Riesz fractional d’Alembertian operator \cite{Riesz:1949, Tarasov:2018},  the Levy fractional parameter, and the Planck mass, respectively. In the limit $\alpha\to 2$, Eq.~\eqref{WDW_frac} reduces to the standard WDW equation.

This fractional extension modifies the quantum spectrum of black holes. In the semiclassical limit, the corresponding mass spectrum is given by
\begin{equation}\label{mass_spectrum_frac}
	M = \left(\frac{\Omega_{d-1}}{\Omega_d}\pi n\right)^{1/d} M_\text{P}, \quad n\in \mathbb{Z}^+,
\end{equation}

where $d=\frac{2}{\alpha}+1$ is the effective fractal dimension of the black hole horizon. Since Levy’s fractional parameter is restricted to the interval $1<\alpha\leq 2$, the corresponding fractal dimension is constrained to the range $2\leq d<3$~\cite{daSilvaJunior:2023, laskin2002fractional}. $\Omega_d$ denotes the volume of a $d$-dimensional unit sphere. The corresponding entropy takes the modified form
\begin{equation}
	S_\text{fractal-BH} = S_\text{BH}^{\,d/2}, \quad S_\text{BH} = 4 \pi G M^2.
\end{equation}
Thus, the entropy no longer scales linearly with the area, but reflects the underlying fractal structure of the horizon. Following the emergent gravity paradigm \cite{Padmanabhan:2012}, the apparent horizon of the universe has the fractal character of black hole horizons. Assuming that cosmological horizons share the same microscopic origin, we extend this construction to the apparent horizon of the Universe. Defining the effective fractal radius $R_\text{eff}$, one obtains
\begin{align}\label{reff_defs}
	& R_\text{eff} = (4\pi)^{\frac{d-2}{4}} \left(\frac{R_H}{L_\text{P}}\right)^{d/2} L_\text{P}, \\ \nonumber
	& A_\text{fractal-H} =  4\pi R_\text{eff}^2, \\ \nonumber
	& V_\text{fractal-H} = \frac{4\pi}{3} R_\text{eff}^3,
\end{align}
where $R_H = 1/H$ denotes the Hubble horizon radius. The modification of the geometric scaling directly impacts the conservation law. The resulting continuity equation is given by $\dot{\rho}_i = -\frac{3d}{2}\,(\rho_i + p_i)H,$ where the redshift $z$ is related to the scale factor by $\left(\frac{a_0}{a(t_{\rm em})}\right)^{d/2} = 1 + z$. Using the holographic equipartition principle, the fractional fractal Friedmann and Raychaudhuri equations for a flat FLRW universe are then obtained as
\begin{align}
	&H^d = \frac{2}{3} (4\pi)^{d/2} L_\text{P}^{4-d} \sum_i \rho_i, \label{friedmann_frac} \\
	&\frac{\ddot a}{a} = -\frac{4\pi G}{3} \left(\frac{2}{3\rho_\text{P}} \sum_j \rho_j \right)^{\frac{2}{d}-1} \sum_i (\rho_i + 3p_i), \label{raychaudhuri_frac}
\end{align}

these equations reduce exactly to the standard Friedmann equations when $d=2$. We consider a cosmic fluid  composed of radiation $\omega=1/3$, pressureless cold matter (baryonic and dark matters) $\omega=0$, and a cosmological constant $\omega_{\Lambda}=-1$. 
From the continuity equation the corresponding energy densities are obtained  by
\begin{equation}\label{continuity}
	\begin{aligned}
		\rho_r(a) &= \rho_{r0}\, a^{-2d}, \\
		\rho_m(a) &= \rho_{m0}\, a^{-3d/2}, \\
		\rho_\Lambda(a) &= \rho_{\Lambda 0}.
	\end{aligned}
\end{equation}
Defining the effective fractal density parameter at present epoch as

\begin{equation}
	\Omega^{(i,frac)}_0=\frac{8\pi G \rho_0}{3H^2_0}(\frac{L_p H_0}{2 \sqrt{\pi}})^{(2-d)},
\end{equation}
the Friedmann equation~\ref{friedmann_frac} can be rewritten as follows
\begin{equation}
	H^2=H_0^2 \left(\Omega_0^{(rad, frac)}(\frac{a}{a_0})^{-2d}+\Omega_0^{(m, frac)}(\frac{a}{a_0})^{\frac{-3d}{2}}+\Omega^{(\Lambda, frac)}\right)^{\frac{2}{d}}.
\end{equation}
The deceleration parameter is then given by
\begin{equation}
	q(a) = 
	\frac{
		2\Omega_0^{(rad, frac)} (\frac{a}{a_0})^{-2d}
		+
		\Omega_0^{(m,frac)} (\frac{a}{a_0})^{-3d/2}
		-
		2 \Omega^{(\Lambda, frac)}
	}{
		2\left[
		\Omega_0^{(rad, frac)} (\frac{a}{a_0})^{-2d}
		+
		\Omega_0^{(m,frac)} (\frac{a}{a_0})^{-3d/2}
		+
		\Omega^{(\Lambda, frac)}
		\right]
	},
\end{equation}
where $\Omega^{(\Lambda, frac)}= \Omega^{(\Lambda, frac)}_0=1-\Omega_0^{(rad, frac)} -\Omega_0^{(m, frac)}$. To determine the value of $d$ and quantify its impact on cosmological evolution, as well as to assess its consistency with the $\Lambda$CDM model and possible deviations from it, we fit the model to observational data.

\section{Observational Datasets and Statistical Methodology}\label{obser}

The aim of our statistical analysis for the EFF cosmological model is to constrain the standard cosmological parameters and the additional model specific parameter by fitting the model to observational data at the background (geometric) and perturbative (growth) levels.
The total likelihood function is defined as
\begin{equation}
	\mathcal{L}_{\rm tot}(p) = \prod_{i} \mathcal{L}_i,
\end{equation}
where the index $i$ runs over the different observational data used in the analysis.
Assuming that the datasets are statistically independent, the total likelihood can be written as
\begin{equation}\label{chi2}
	\mathcal{L}_{\rm tot} \simeq \exp\left(-\frac{\chi^2_{\rm tot}}{2}\right),
\end{equation}
with the total chi-square given by
\begin{equation}\label{eq:chi2_tot}
	\chi^2_{\rm tot}(p) = \sum_i \chi^2_i(p).
\end{equation}
To investigate the impact of different observational probes, we consider three dataset combinations:
\begin{itemize}
	\item late--time (LT) probes: Type Ia supernovae (Pantheon+SH0ES), direct measurements of the Hubble parameter $H(z)$, and measurements of the growth rate of structure through $f\sigma_8(z)$.
	
	\item LT+DESI DR2 BAO+BBN: the LT dataset combined with DESI DR2 BAO measurements, and the Big Bang Nucleosynthesis (BBN) constraint on $\Omega_b h^2$.
	
	\item LT+DESI DR2 BAO+CMB: the LT dataset combined with DESI DR2 BAO measurements, and the Planck 2018 CMB distance priors.
\end{itemize}
The parameter vector $p$ depends on the cosmological model under consideration. For the $\Lambda$CDM model, we adopt
$p = (\Omega_{b0}, \Omega_{\mathrm{dm}0}, H_0, \sigma_8, M_B),$
where $\Omega_{b0}$ and $\Omega_{\mathrm{dm}0}$ are the present day baryon and cold dark matter density parameters, $H_0$ is the Hubble constant, $\sigma_8$ is the amplitude of matter fluctuations, and $M_B$ is the absolute magnitude of Type Ia supernovae. For the EFF model, the parameter space is extended to
$p = (\Omega_{b0}, \Omega_{\mathrm{dm}0}, H_0, d, \sigma_8, M_B),$ where $d$ is the effective fractal dimension, representing deviations from standard cosmology.
The parameter estimation is performed using the Markov Chain Monte Carlo (MCMC) technique based on the Metropolis--Hastings algorithm, implemented through the \texttt{emcee} ensemble sampler \cite{foreman2013emcee}.

We use broad uniform priors for the model parameters, as summarized in Table~\ref{tab:priors}. The parameter $d$ is restricted to $d \ge 2$ to ensure the physical consistency of the EFF model. For the LT dataset, we additionally impose a Gaussian prior, $\Omega_{b0}h^2 \sim \mathscr{N}(0.022, 0.010^2)$, to prevent the chains from exploring unphysical regions of parameter space. The convergence of the MCMC chains is assessed using the Gelman--Rubin diagnostic~\cite{Gelman:1992zz}. The sampling configuration and convergence statistics are summarized in Table~\ref{tab:mcmc_settings}, including the number of chains ($N_{\rm ch}$), walkers ($N_{\rm w}$), total steps per walker, burn-in length, thinning, the maximum autocorrelation time ($\tau_{\max}$), the maximum value of $\hat{R}-1$, and the minimum effective sample size (ESS). In all cases, the chains satisfy our convergence criteria, with $\max(\hat{R}-1) \le 2\times10^{-4}$ and $\mathrm{ESS}_{\min} > 1.3\times10^4$. Posterior distributions and credible intervals are derived using the \texttt{GetDist} package~\cite{Lewis:2019}.

\begin{table}[htbp]
\centering
\caption{The prior distributions for the parameters used in the MCMC analysis.}
\label{tab:priors}
\begin{tabular}{lc}
\hline\hline
Parameter & Prior \\
\hline
$\Omega_{dm0}$      & $\mathscr{U}(0, 1)$ \\
$\Omega_{b0}$       & $\mathscr{U}(0, 1)$ \\
$H_0$ [km/s/Mpc]    & $\mathscr{U}(40, 100)$ \\
$d$                 & $\mathscr{U}(2, 3)$ \\
$\sigma_8$          & $\mathscr{U}(0, 2)$ \\
$M_b$               & $\mathscr{U}(-20.5, -18.0)$ \\
$\Omega_{b0}h^2$ (LT only) & $\mathscr{N}(0.022, 0.010^2)$ \\
\hline\hline
\end{tabular}
\end{table}

\begin{table}[htbp]
\centering
\scriptsize
\caption{MCMC sampling configuration and convergence diagnostics for the EFF model.}
\label{tab:mcmc_settings}
\resizebox{\linewidth}{!}{%
\begin{tabular}{lcccccccc}
\hline\hline
Dataset & $N_{\rm ch}$ & $N_{\rm w}$ & Steps & Burn-in & Thin & $\tau_{\max}$ & max($\hat{R}-1$) & Min ESS \\
\hline
LT                   & 2 & 80 & 20000 & 5000 & 2 & 42.29 & 0.0001 & 14187 \\
LT+DESI BAO+BBN      & 2 & 80 & 20000 & 5000 & 2 & 42.03 & 0.0002 & 14275 \\
LT+DESI BAO+CMB      & 2 & 80 & 20000 & 5000 & 2 & 43.31 & 0.0001 & 13855 \\
\hline\hline
\end{tabular}%
}
\end{table}

In the following subsections, we briefly describe each dataset used in this work.

\subsection{Cosmic Microwave Background (CMB)}
The position of the acoustic peaks of the CMB can be described by the CMB distance priors, defined by the parameter set $(R, \ell_A, \omega_b \equiv \Omega_b h^2)$~\cite{chen:2019}. These quantities provide a compressed description of the CMB data and can be used to constrain cosmological models.
The inverse covariance matrix of these distance priors is taken from the Planck 2018 baseline $\Lambda$CDM analysis, and is given by
\begin{equation}
	\mathrm{Cov}^{-1}_{CMB} =
	\begin{pmatrix}
		9.320529\times10^{4} & -1.344780\times10^{3} & 1.63416998\times10^{6} \\
		-1.344780\times10^{3} & 1.605600\times10^{2} & 5.058240\times10^{3} \\
		1.63416998\times10^{6} & 5.058240\times10^{3} & 7.89057112\times10^{7}
	\end{pmatrix}.
\end{equation}

\noindent
The shift parameter is defined as
\begin{equation}
	R(z_*)=\frac{(1+z_*)D_A(z_*)\sqrt{\Omega_m H^2_0}}{c},
\end{equation}
and the acoustic scale is given by
\begin{equation}\label{eq38}
	l_A=(1+z_*)\frac{\pi D_A(z_*)}{r_s(z_*)},
\end{equation}
where $r_s(z_*)$ is the comoving sound horizon at the decoupling epoch. $D_A$ is the angular diameter distance.
The reference values of the distance priors from \textit{Planck} 2018 are
$(
R_\text{obs} = 1.7502, l_{A,\text{obs}} = 301.471, \omega_{b,\text{obs}} = 0.02236)
$~\cite{chen:2019}.
The chi-square of CMB is given by
\begin{equation}
	\chi^2_{\rm CMB}
	= {x}^{\mathrm{T}}
	\,\mathrm{Cov}^{-1}_{\rm CMB}\,
	{x}.
\end{equation}

\noindent
which ${x} = \left( R_{\mathrm{th}} - R_{\mathrm{obs}},\;
\ell_{A,\mathrm{th}} - \ell_{A,\mathrm{obs}},\;
\omega_{b,\mathrm{th}} - \omega_{b,\mathrm{obs}} \right)^{T}$.

\subsection{Type Ia supernovae (Pantheon+SH0ES)}
Type Ia supernovae (SNe Ia) serve as standardizable candles, enabling a precise reconstruction of the cosmic expansion history through the distance modulus redshift relation. In this analysis, we use the \textit{Pantheon+SH0ES} compilation~\cite{Scolnic:2022}, which consists of 1701 light-curve measurements from 1550 spectroscopically confirmed SNe~Ia, together with 72 Cepheid calibrators. The combined dataset, compiled from 18 independent surveys, spans the redshift range $0.001 < z < 2.26$.

The theoretical distance modulus is defined as
\begin{equation}
	\mu_{\mathrm{th}}(z) = 5 \log_{10} \left[ \frac{d_L(z)}{\mathrm{Mpc}} \right] + 25,
\end{equation}
where the luminosity distance is given by
\begin{equation}
	d_L(z) = (1+z)\, c \int_{0}^{z} \frac{dz'}{H(z')}.
\end{equation}
The residual vector for supernovae is defined as
\begin{equation}
	Q_{\mathrm{SN}} = 
	m_B^{\mathrm{SN}} - M_B - \mu_{\mathrm{th}}(z),
\end{equation}
and the corresponding chi-square is
\begin{equation}
	\chi^2_{\mathrm{SN}} =
	Q_{\mathrm{SN}}^{\mathrm{T}}
	\, C_{\mathrm{SN}}^{-1}\,
	Q_{\mathrm{SN}}.
\end{equation}
For Cepheid calibrators, we define
\begin{equation}
	Q_{\mathrm{Ceph}}
	= m_B^{\mathrm{Ceph}} - M_B - \mu_{\mathrm{Ceph}},
\end{equation}
with the associated chi-square
\begin{equation}
	\chi^2_{\mathrm{Ceph}} =
	Q_{\mathrm{Ceph}}^{\mathrm{T}}
	\, C_{\mathrm{Ceph}}^{-1}\,
	Q_{\mathrm{Ceph}}.
\end{equation}
Here, $C_{\mathrm{SN}}$ and $C_{\mathrm{Ceph}}$ denote the covariance matrices of the Pantheon+SH0ES compilation corresponding to the supernova and Cepheid samples, respectively. The combined chi-square for the Pantheon+SH0ES dataset is then given by
\begin{equation}
	\chi^2_{\mathrm{SN+Ceph}}
	= \chi^2_{\mathrm{SN}} + \chi^2_{\mathrm{Ceph}}.
\end{equation}

\subsection{Baryon Acoustic Oscillations (BAO DESI DR2)}

Baryon Acoustic Oscillations (BAO) provide a well-calibrated comoving standard ruler determined by the sound horizon at the baryon drag epoch, $r_{\rm d}$. Measurements of the BAO feature in the transverse and radial directions constrain the quantities
\begin{equation}
	\frac{D_M(z)}{r_{\rm d}},
	\qquad
	\frac{D_H(z)}{r_{\rm d}}
	=
	\frac{c}{H(z)\,r_{\rm d}},
\end{equation}
as well as the isotropic combination
\begin{equation}
	\frac{D_V(z)}{r_{\rm d}}
	=
	\frac{1}{r_{\rm d}}
	\left[
	z\,D_H(z)\,D_M^2(z)
	\right]^{1/3}.
\end{equation}
Here, $D_H(z)=c/H(z)$ is the Hubble distance, while $D_M(z)$ denotes the transverse comoving distance,
\begin{equation}
	D_M(z)=S_k\!\left(\chi(z)\right),
\end{equation}
with
\begin{equation}
	\chi(z)=\frac{c}{H_0}\int_0^z \frac{dz'}{E(z')},
	\qquad
	E(z)=\frac{H(z)}{H_0}.
\end{equation}
The curvature-dependent function $S_k(\chi)$ is defined as
\begin{equation}
	S_k(\chi)=
	\begin{cases}
		\dfrac{1}{\sqrt{\Omega_{k0}}}
		\sinh\!\left(\sqrt{\Omega_{k0}}\,\chi\right),
		& \Omega_{k0}>0, \\[0.3cm]
		\chi,
		& \Omega_{k0}=0, \\[0.3cm]
		\dfrac{1}{\sqrt{-\Omega_{k0}}}
		\sin\!\left(\sqrt{-\Omega_{k0}}\,\chi\right),
		& \Omega_{k0}<0 .
	\end{cases}
\end{equation}
We use the DESI Data Release 2 (DR2) BAO measurements~\cite{DESI:2025}. The sample consists of post-reconstruction clustering measurements from more than $14$ million galaxies and quasars, yielding $13$ BAO distance constraints spanning the redshift range $0.295<z<2.33$. In particular, the sample contains measurements from the BGS, LRG, ELG, QSO, and Ly$\alpha$ forest tracers.

The comoving sound horizon at the baryon drag epoch, $r_{\rm d}$, is evaluated within the assumed background cosmology, adopting $N_{\rm eff}=3.046$ and the same radiation content adopted in the CMB analysis.
Assuming a Gaussian likelihood, the BAO chi-square function is defined as
\begin{equation}
	\chi^2_{\mathrm{BAO}}
	=
	(d_{\rm obs}-d_{\rm th})^{T}
	C^{-1}_{\mathrm{BAO}}
	(d_{\rm obs}-d_{\rm th}),
\end{equation}
where $d_{\rm obs}$ represents the vector of the $13$ BAO measurements from DESI DR2, consisting of $D_V/r_{\rm d}$, $D_M/r_{\rm d}$, or $D_H/r_{\rm d}$ according to the redshift interval, while $d_{\rm th}$ corresponds to the associated theoretical predictions. The matrix $C_{\mathrm{BAO}}$ denotes the full covariance matrix released by the DESI collaboration for the combined galaxy and quasar sample.

\subsection{Observational Hubble Data (OHD)}
In addition, we include 40 measurements of the Hubble parameter $H(z)$ obtained from cosmic chronometer observations and radial BAO clustering analyses~\cite{Tamri:2026, Mohebi:2026}.
The corresponding $\chi^2$ is defined as
\begin{equation}
	\chi^2_{H(z)} = 
	\sum_{i} \frac{\left[ H_{\rm th}(z_i) - H_{\rm obs}(z_i) \right]^2}{\sigma_{H,i}^2} ,
\end{equation}
where $H_{\rm th}(z_i), H_{\rm obs}(z_i),\sigma_{H,i} $ are the theoretical value, the observed value, and the reported uncertainty of each data point, respectively.
\begin{table}[t]
	\centering
	\caption{List of the 40 $H(z)$ data points used in this analysis~\cite{Tamri:2026,Mohebi:2026}.}
	\label{tab:Hz_data}
	\begin{tabular}{ccc @{\hspace{1.2cm}} ccc}				
		\toprule[0.4mm]
		$z$ & $H(z)$ & $\sigma_H$ & $z$ & $H(z)$ & $\sigma_H$ \\
		\midrule
		0.070 & 69.00 & 19.60  & 0.480 & 97.00 & 62.00 \\
		0.090 & 69.00 & 12.00  & 0.500 & 72.10 & 34.60 \\
		0.100 & 69.00 & 12.00  & 0.570 & 89.20 &  3.60 \\
		0.120 & 68.60 & 26.20  & 0.593 & 104.00 & 13.00 \\
		0.170 & 83.00 &  8.00  & 0.680 & 92.00 &  8.00 \\
		0.180 & 75.00 &  4.00  & 0.730 & 97.30 &  7.00 \\
		0.200 & 72.90 & 29.60  & 0.750 & 105.00 & 10.80 \\
		0.240 & 79.69 &  2.65  & 0.753 & 98.80 & 33.60 \\
		0.270 & 77.00 & 14.00  & 0.800 & 113.10 & 32.50 \\
		0.280 & 88.80 & 36.60  & 0.875 & 125.00 & 17.00 \\
		0.350 & 82.10 &  4.80  & 0.900 & 117.00 & 23.00 \\
		0.352 & 83.00 & 14.00  & 1.037 & 154.00 & 20.00 \\
		0.380 & 83.00 & 13.50  & 1.260 & 135.00 & 65.00 \\
		0.400 & 77.00 & 10.20  & 1.300 & 168.00 & 17.00 \\
		0.425 & 87.10 & 11.20  & 1.363 & 160.00 & 33.60 \\
		0.429 & 91.80 &  5.30  & 1.430 & 177.00 & 18.00 \\
		0.440 & 82.60 &  7.80  & 1.530 & 140.00 & 14.00 \\
		0.450 & 92.80 & 12.90  & 1.750 & 202.00 & 40.00 \\
		0.470 & 89.00 & 34.00  & 1.965 & 186.50 & 50.40 \\
		0.478 & 80.90 &  9.00  & 2.300 & 224.00 &  8.00 \\
		\bottomrule
	\end{tabular}
\end{table}

\subsection{Big Bang Nucleosynthesis (BBN)}

BBN provides an additional and independent constraint on cosmological models through the primordial abundances of light elements produced in the early Universe. To incorporate this information, we make use of the public \texttt{PRIMAT} BBN framework~\cite{pitrou2018precision}, adopting the most recent grid release. The calculation includes quantum electrodynamics corrections, as well as the effects of incomplete neutrino
decoupling. In this setup, $\Delta N = 0$ corresponds to an effective number of relativistic species $N_{\rm eff}\simeq 3.044$. In our analysis we keep $N_{\rm eff}$ fixed, consistent with the standard value used in the
Planck 2018 analyses.

The predicted primordial helium mass fraction $Y_p^{\rm th}$ is obtained by interpolating the PRIMAT grid as a function of the baryon density
$\omega_b \equiv \Omega_{b0}h^2$, evaluated at $\Delta N=0$. This prediction is then confronted with the observational estimate
$Y_p^{\rm obs}=0.2448 \pm 0.0033$~\cite{aver2022comprehensive}. Assuming Gaussian uncertainties, the likelihood contribution from BBN can be written as
\begin{equation}
	\chi^2_{\rm BBN} =
	\left(
	\frac{Y_p^{\rm th}(\Omega_{b0}h^2) - Y_p^{\rm obs}}
	{\sigma_{Y_p}}
	\right)^2 ,
\end{equation}
with an associated uncertainty of $\sigma_{Y_p}=0.0033$.

In the present analysis, the BBN constraint is included as an external likelihood based on the standard \texttt{PRIMAT} interpolation grid at fixed $\Delta N=0$. The theoretical helium abundance $Y_p^{\rm th}$ is obtained by interpolating the grid at the EFF-inferred baryon density, $\omega_b=\Omega_{b0}h^2$. However, the modified EFF background evolution is not used in a fully self-consistent recomputation of the BBN process.

\subsection{Structure Formation}

To investigate the evolution of cosmic structures, we include observational datasets containing measurements of $f\sigma_8(z)$. This observable traces the growth rate of matter perturbations and is directly related to the matter density contrast $\delta_m$. The quantity $\sigma_8(z)$ represents the redshift-dependent root mean square fluctuation of the linear density field within spheres of radius $R = 8h^{-1}\,\mathrm{Mpc}$, and is given by
\begin{equation}
	\sigma_8(a)=\sigma_8(a=1)\,
	\frac{\delta_m(a)}{\delta_m(a=1)}.
\end{equation}

\noindent
The evolution of linear matter density perturbations in a matter-dominated Universe is governed by~\cite{Mehrabi:2015kta}
\begin{equation}
	\ddot{\delta}_m + 2H\,\dot{\delta}_m
	-4\pi G\,\rho_m(t)\,\delta_m = 0.
\end{equation}

\noindent
Using the relation
$
\frac{ d}{ dt}=aH\frac{ d}{da}
$,
together with $\rho_m(a)=\rho_{m0} a^{-3d/2}$ and 
\begin{equation}
	\Omega^{(m, \mathrm{frac})}_{0}
	=
	\frac{
		\frac{2}{3}(4\pi)^{d/2}
		L_p^{4-d} \rho_{m0}
	}{
		H_0^d
	},
\end{equation}
we obtain the modified growth equation
\begin{equation}
	\delta_m''(a)
	+
	\left[
	\frac{3}{a}
	+
	\frac{H'(a)}{H(a)}
	\right]
	\delta_m'(a)
	-
	\frac{3}{2}\,
	(4\pi)^{1-d/2}\,
	L_p^{\,d-2}\,
	\Omega^{(m,\mathrm{frac})}_{0}\,
	\frac{H_0^d}{H^2(a)}\,
	a^{-(3d/2+2)}\,
	\delta_m(a)
	= 0,
	\label{eq:delta_m_evolution}
\end{equation}
where primes denote derivatives with respect to the scale factor.

\noindent
We set the initial scale factor to $a_i = 10^{-3}$, corresponding to the matter-dominated epoch. In this limit, the growing mode solution behaves as $\delta_m(a)\propto a^{3d/4}$. Accordingly, we set $\delta_m(a_i)=1.6\times10^{-4}$ and impose the consistent initial condition~\cite{Batista:2013oca}
\begin{equation}
	\delta_m'(a_i)
	= \frac{3d}{4}\,\frac{\delta_m(a_i)}{a_i}.
\end{equation}

\noindent
We then numerically solve Eq.~\eqref{eq:delta_m_evolution} from this initial state up to the present epoch $a_0 = 1$, allowing us to compute the growth rate $f(a) = d\ln \delta_m / d\ln a$ and the observable combination $f\sigma_8(a)$. In this analysis, we employ 26 independent observational measurements of $f(z)\sigma_8(z)$~\cite{Mohebi:2026} listed in Table~\ref{tab:fsigma8}.
\begin{table}[htbp]
	\centering
	\caption{Measurements of the growth rate quantity $f\sigma_8$ at different redshifts~\cite{Mohebi:2026}.}
	\label{tab:fsigma8}
	\begin{tabular}{ccc @{\hspace{1.2cm}} ccc}
		\hline
		$z$ & $f\sigma_8$ & $\sigma$ & $z$ & $f\sigma_8$ & $\sigma$ \\
		\hline
		0.02  & 0.428 & 0.0465 & 0.60  & 0.550 & 0.120 \\
		0.025 & 0.390 & 0.110  & 0.70  & 0.473 & 0.041 \\
		0.067 & 0.423 & 0.055  & 0.71  & 0.484 & 0.055 \\
		0.10  & 0.370 & 0.130  & 0.73  & 0.437 & 0.072 \\
		0.15  & 0.530 & 0.160  & 0.85  & 0.315 & 0.095 \\
		0.17  & 0.510 & 0.060  & 0.86  & 0.400 & 0.110 \\
		0.18  & 0.360 & 0.090  & 0.92  & 0.422 & 0.046 \\
		0.295 & 0.378 & 0.094  & 1.05  & 0.280 & 0.080 \\
		0.38  & 0.440 & 0.060  & 1.32  & 0.375 & 0.039 \\
		0.38  & 0.497 & 0.045  & 1.40  & 0.482 & 0.116 \\
		0.44  & 0.413 & 0.080  & 1.48  & 0.462 & 0.045 \\
		0.51  & 0.516 & 0.062  & 1.49  & 0.435 & 0.045 \\
		0.60  & 0.390 & 0.063  & 1.944 & 0.364 & 0.106 \\
		\hline
	\end{tabular}
\end{table}

\section{Numerical Results from MCMC}\label{num}

In this section, we use the MCMC method to estimate the best-fit values of the free parameters of the EFF model. The constraints are derived from 40 $H(z)$ measurements, the 1701-point Pantheon+SH0ES supernova compilation, 26 $f\sigma_8(z)$ measurements, 13 DESI DR2 BAO data points, a BBN constraint on $Y_p$ based on the PRIMAT grid, and 3 CMB distance priors. 

To examine the consistency of our results and the impact of different cosmological probes, we consider three combinations of observational data. The LT dataset includes only late-time probes, the Pantheon+SH0ES supernova sample, $H(z)$ measurements, and $f\sigma_8(z)$ data. In this case, a weak Gaussian prior on $\Omega_{b0}h^2$ is also imposed, as listed in Table~\ref{tab:priors}. The combination LT + DESI DR2 BAO + BBN extends the late-time dataset by including 13 DESI DR2 BAO measurements and a BBN constraint for the primordial helium abundance $Y_p$, constructed from a PRIMAT-based interpolation grid with 1300 points spanning 52 values of $\Omega_b h^2$ and 25 values of $\Delta N$. 

Finally, LT + DESI DR2 BAO + CMB replaces the BBN information with 3 compressed CMB distance priors from recombination, while retaining the same 13 DESI DR2 BAO measurements. This setup allows us to assess how early-Universe information influences the inferred values of $H_0$, $\sigma_8$, and the effective fractal dimension $d$.

The best-fit values of the parameters for the flat EFF model, along with their $68\%$ confidence level uncertainties, are summarized in Table~\ref{tab:results_flat_EFF}. An important result is that the present dark matter density parameter remains clearly non-zero for all dataset combinations. In particular, the combination including CMB distance priors yields 
$\Omega_{dm0}=0.2722\pm0.0036$, 
which is fully consistent with the value obtained in the standard $\Lambda$CDM model. 
This indicates that, despite the modified cosmological dynamics introduced by the fractal parameter $d$, the observational data still strongly favor the presence of a non-baryonic dark matter component. Furthermore, while the LT and LT + DESI DR2 BAO + BBN tend to prefer a higher value of the Hubble constant, $H_0 \approx 71$~km\,s$^{-1}$\,Mpc$^{-1}$, the inclusion of CMB information (LT + DESI DR2 BAO + CMB) pulls the posterior distribution back toward the standard Planck-like value, $H_0 \approx 67.3$~km\,s$^{-1}$\,Mpc$^{-1}$. Consequently, in terms of the $H_0$ tension, the EFF model exhibits a behavior similar to that of the standard $\Lambda$CDM model.

\begin{figure}[H]
	\centering
	\includegraphics[width=0.7\columnwidth]{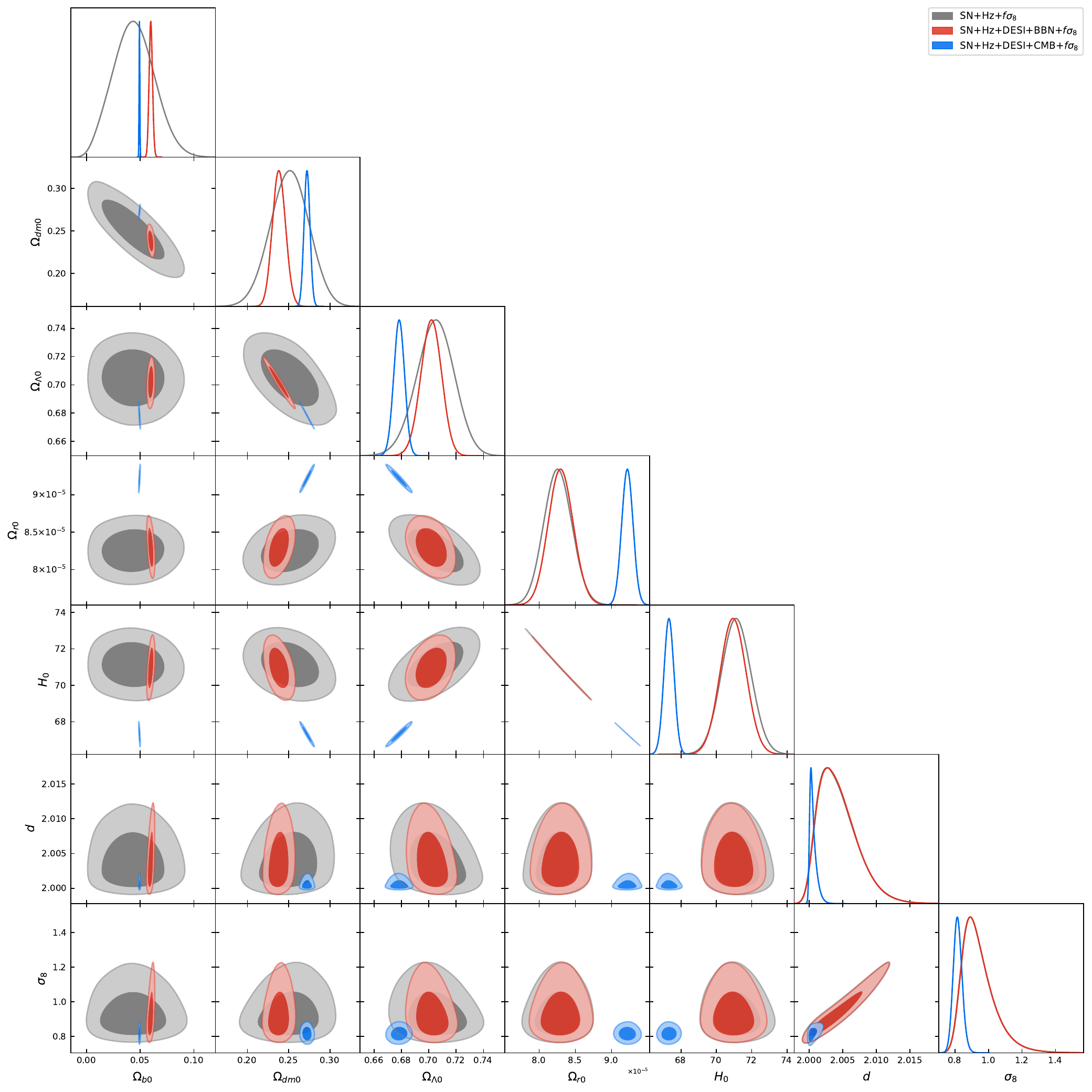}
	\caption{
		The posterior distributions and corresponding $1\sigma$ and $2\sigma$ confidence contours of the main EFF model parameters obtained from the three dataset combinations: LT (gray), LT+DESI DR2 BAO+BBN (red), and LT+DESI DR2 BAO+CMB (blue).}
	\label{fig:EFF}
\end{figure}

Figure~\ref{fig:EFF} shows the marginalized posterior distributions with the $68\%$ and $95\%$ confidence contours for the EFF model parameters. The figure compares the constraints obtained from three dataset combinations: LT (grey), LT+DESI DR2 BAO+BBN (red), and LT+DESI DR2 BAO+CMB (blue). The parameters shown include the present-day density parameters ($\Omega_{b0}$, $\Omega_{dm0}$, $\Omega_{\Lambda0}$, and $\Omega_{r0}$), the Hubble constant $H_0$, the effective fractal dimension parameter $d$, and the amplitude of matter fluctuations $\sigma_8$.

The LT and LT+BAO+BBN datasets allow mild deviations from the $\Lambda$CDM limit $d=2$. However, once the CMB distance priors are included, the posterior distribution sharply collapses around
$2.0004^{+0.0006}_{-0.0003}$,
indicating that any fractal deviation from standard cosmology must be extremely small. 

The LT contours exhibit a clear correlation in the $(d,H_0)$ plane. Larger values of $d$ allow a higher value of $H_0$. This degeneracy is progressively reduced when BAO+BBN dataset is added and is essentially eliminated once the CMB data are included, which fixes the sound horizon scale and constrains $H_0$ toward the value preferred by the $\Lambda$CDM model.

The parameter $\sigma_8$ shows a mild degeneracy with $d$ in the LT dataset because the modified continuity equation alters the redshift scaling of matter and therefore affects the growth of cosmic structures. However, when early-time data are incorporated, the inferred value of $\sigma_8$ shifts toward the region preferred by $\Lambda$CDM, indicating that growth data alone cannot support significant deviations in $d$ once early-Universe constraints are imposed.
\begin{table}[t]
	\centering
	\footnotesize
	\caption{\small Constraints on the cosmological parameters of the flat EFF model from different dataset combinations, with uncertainties at the $68\%$ credible level from the marginalized posteriors.
}
	\setlength{\tabcolsep}{12pt}
	\renewcommand{\arraystretch}{1.5}
	\resizebox{\textwidth}{!}{%
		\begin{tabular}{lccc}
			\toprule
			Parameter & LT & LT+BAO+BBN & LT+BAO+CMB \\
			\midrule
			$\Omega_{b_0}$ & $0.0437^{+0.0195}_{-0.0190}$ & $0.0599^{+0.0016}_{-0.0015}$ & $0.0493^{+0.0003}_{-0.0003}$ \\
$\Omega_{dm_0}$ & $0.2513^{+0.0230}_{-0.0232}$ & $0.2383^{+0.0080}_{-0.0078}$ & $0.2722^{+0.0036}_{-0.0036}$ \\
$\Omega_{\Lambda_0}$ & $0.7049^{+0.0133}_{-0.0137}$ & $0.7018^{+0.0074}_{-0.0076}$ & $0.6783^{+0.0039}_{-0.0039}$ \\
$\Omega_{r_0}$ & $(8.2579^{+0.1935}_{-0.1903}) \times 10^{-5}$ & $(8.2981^{+0.1722}_{-0.1687}) \times 10^{-5}$ & $(9.2175^{+0.0771}_{-0.0767}) \times 10^{-5}$ \\
$H_0$ & $71.12^{+0.83}_{-0.82}$ & $70.95^{+0.73}_{-0.72}$ & $67.32^{+0.28}_{-0.28}$ \\
$d$ & $2.0038^{+0.0033}_{-0.0024}$ & $2.0039^{+0.0033}_{-0.0024}$ & $2.0004^{+0.0006}_{-0.0003}$ \\
$\sigma_8$ & $0.9212^{+0.1105}_{-0.0708}$ & $0.9206^{+0.1097}_{-0.0713}$ & $0.8156^{+0.0267}_{-0.0249}$ \\
$M_B$& $-19.329 \pm 0.023$ & $-19.33341 \pm 0.021$ & $-19.43803 \pm 0.008$ \\
			\midrule
			$\chi^2_{\min}$ & $1570.99688$ & $1582.68787$ & $1632.92053$ \\
			$\rm AIC$ & $1582.99688$ & $1594.68787$ & $1644.92053$ \\
$\rm BIC$ & $1615.85911$ & $1627.59745$ & $1677.89685$ \\
			\bottomrule
		\end{tabular}
	}
	\label{tab:results_flat_EFF}
\end{table}

\begin{figure}[H]
	\centering
	\includegraphics[width=0.7\columnwidth]{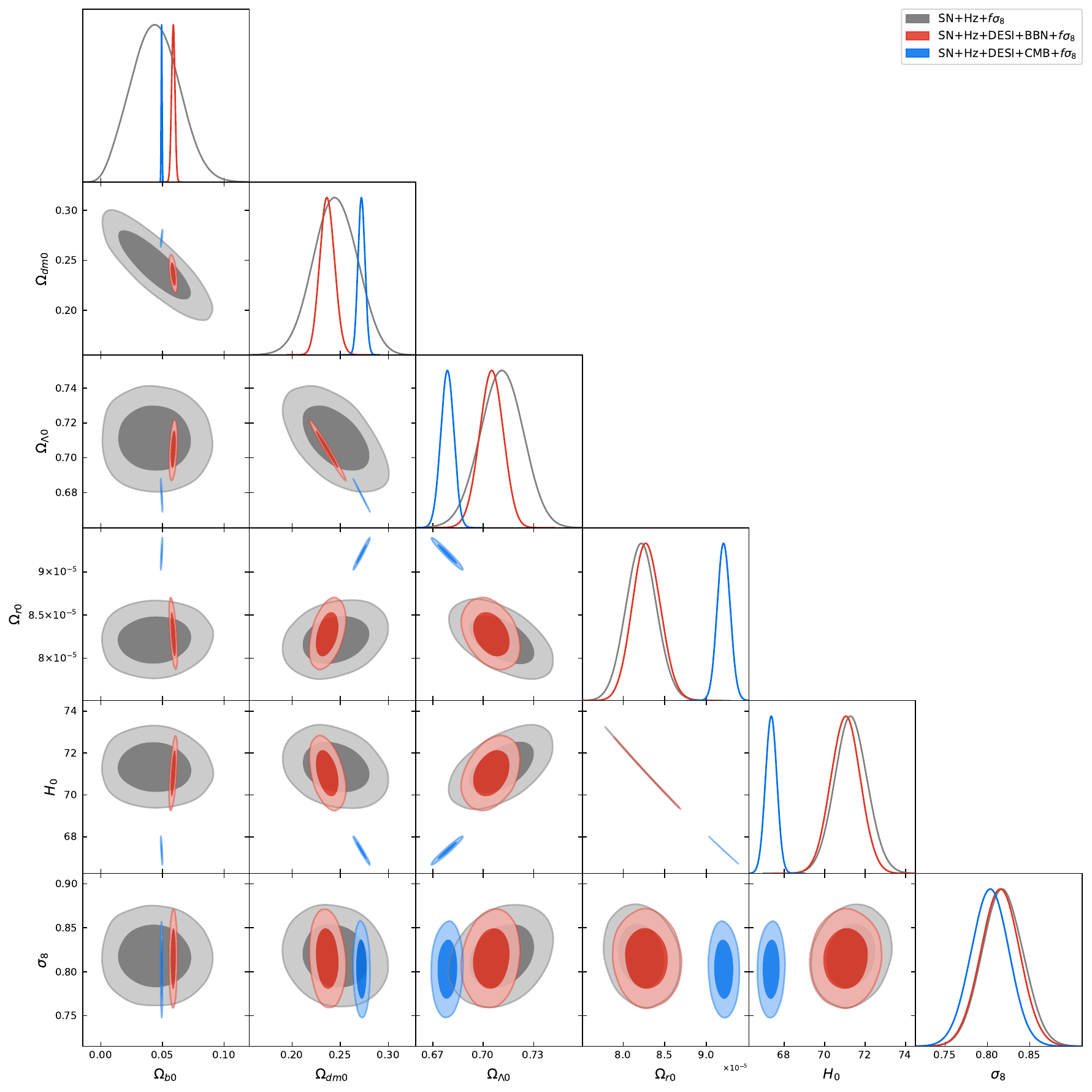}
	\caption{The posterior distributions and corresponding $1\sigma$ and $2\sigma$ confidence contours of the main $\Lambda$CDM model parameters obtained from the three dataset combinations: LT (gray), LT+DESI DR2 BAO+BBN (red), and LT+DESI DR2 BAO+CMB (blue).}
	\label{fig:lcdm}
\end{figure}

Figure~\ref{fig:lcdm} compares the cosmological constraints for the $\Lambda$CDM model derived from three dataset combinations including LT (grey), LT+DESI DR2 BAO+BBN (red), and LT+DESI DR2 BAO+CMB (blue). The inclusion of the additional parameter $d$ in the EFF model broadens several degeneracy directions in the LT case. 
However, once BAO+BBN and especially CMB priors are added, the allowed region collapses to a narrow band centered on the standard $\Lambda$CDM parameters.  
This behavior confirms that current observations tightly constrain fractional deviations in the EFF framework.

Table~\ref{tab:results_flat_LCDM} summarizes the best-fit values of the cosmological parameters for the flat $\Lambda$CDM model together with their corresponding $68\%$ confidence level uncertainties.
\begin{table*}[t]
	\centering
	\footnotesize
	\caption{\small Constraints on the cosmological parameters of the flat $\Lambda$CDM model from different dataset combinations, with uncertainties at the $68\%$ credible level from the marginalized posteriors.}
	\setlength{\tabcolsep}{9pt}
	\renewcommand{\arraystretch}{1.5}
	\resizebox{\textwidth}{!}{%
		\begin{tabular}{lccc}
			\toprule
			Parameter & LT & LT+BAO+BBN & LT+BAO+CMB \\
			\midrule
			$\Omega_{b0}$ & $0.0440^{+0.0195}_{-0.0193}$ & $0.0588^{+0.0013}_{-0.0014}$ & $0.0494\pm0.0003$ \\
			
			$\Omega_{dm0}$ & $0.2447^{+0.0229}_{-0.0225}$ & $0.2363^{+0.0078}_{-0.0075}$ & $0.2719\pm0.0036$ \\
			
			$\Omega_{\Lambda0}$ & $0.7110^{+0.0126}_{-0.0128}$ & $0.7048^{+0.0070}_{-0.0072}$ & $0.6786\pm0.0039$ \\
			
			$\Omega_{r0}$ & $(8.2172^{+0.1855}_{-0.1842})\times10^{-5}$ & $(8.2776^{+0.1727}_{-0.1651})\times10^{-5}$ & $(9.2088\pm0.076)\times10^{-5}$ \\
			
			$H_0$ & $71.30^{+0.81}_{-0.79}$ & $71.04^{+0.72}_{-0.73}$ & $67.35\pm0.28$ \\
			
			$\sigma_8$ & $0.8177^{+0.0243}_{-0.0237}$ & $0.8150^{+0.0227}_{-0.0229}$ & $0.8030^{+0.0223}_{-0.0224}$ \\
			$M_B$ & $-19.32572 \pm 0.022$ & $-19.33168 \pm 0.021$ & $-19.43710 \pm 0.008$ \\
			\midrule
			$\chi^2_{\min}$ & $1571.53132$ & $	1583.33913$ & $1632.92754$ \\
			AIC & $1581.53132$ & $1593.33913$ & $1642.92754$ \\
			BIC & $1608.91651$ & $1620.76378$ & $1670.35780$ \\
			\bottomrule
		\end{tabular}
	}
	\label{tab:results_flat_LCDM}
\end{table*}
The comparison of Figure~\ref{fig:EFF} and Figure~\ref{fig:lcdm} indicates that the EFF model fits the data comparably well to $\Lambda$CDM, but leads to slightly different parameter correlations due to the presence of the additional parameter $d$. This extra degree of freedom introduces wider degeneracy regions among parameters such as $H_0$, $\sigma_8$, and the matter density, particularly for the late-time dataset, while the inclusion of early-Universe probes significantly reduces these degeneracies and brings the constraints closer to those of $\Lambda$CDM.

For further statistical analysis, we compute the Akaike Information Criterion (AIC) and the Bayesian Information Criterion (BIC):
\begin{equation}
	\begin{aligned}
		\mathrm{AIC} &= \chi^2_{\rm min} + 2k,\\
		\mathrm{BIC} &= \chi^2_{\rm min} + k \ln N,
	\end{aligned}
\end{equation}
where $k$ is the number of free parameters, which is $k=5$ for the flat $\Lambda$CDM model and $k=6$ for the flat EFF model. $N$ is the number of data points which depends on the dataset combination. $N=1767$ for the LT dataset, $N=1781$ for the LT+BAO+BBN combination, and $N=1783$ for the LT+BAO+CMB combination. $\chi^2_{\rm min}$ is the minimum chi-square. 
From Table~\ref{tab:results_flat_EFF} and Table~\ref{tab:results_flat_LCDM}, the differences with respect to the reference $\Lambda$CDM model are 
$\Delta \mathrm{AIC} \approx \{1.47, 1.35, 2.00\}$ 
and 
$\Delta \mathrm{BIC} \approx \{6.94, 6.83, 7.54\}$ 
for the LT, LT+BAO+BBN, and LT+BAO+CMB dataset combinations, respectively. 

In particular, the $\Delta\mathrm{AIC}$ values remain close to or below $2$, indicating that both models receive comparable statistical support from the observational data according to the AIC. 
On the other hand, the $\Delta\mathrm{BIC}$ values lie in the range $\sim 6.8$--$7.5$. According to the Jeffreys' scale, this range provides strong evidence in favor of the simpler $\Lambda$CDM model over the EFF model, primarily due to the penalty associated with the additional free parameter in the latter.

At the end of this section, using the best-fit values, we examine the redshift evolution of the cosmological parameters and show the corresponding plots.
\begin{figure}[H]
	\centering
	\includegraphics[width=\columnwidth]{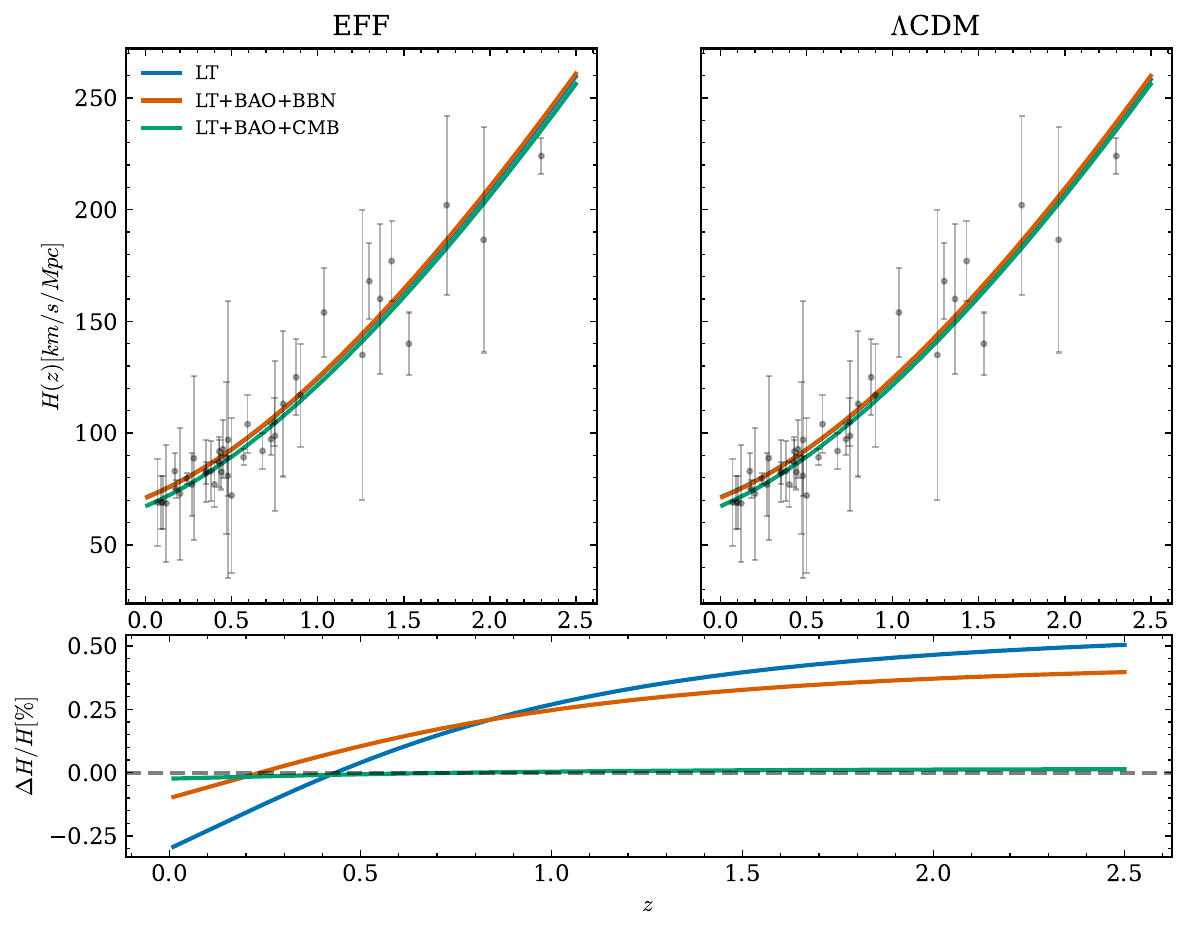}
	\caption{Hubble parameter $H(z)$ as a function of redshift for the $\Lambda$CDM (right panel) and EFF (left panel) models, compared with observational data points for the three dataset combinations. The lower panel shows the relative difference.}
	\label{fig:Hz}
\end{figure}
In the Figure~\ref{fig:Hz}, we have shown the Hubble parameter $H(z)$ as a function of redshift for both the EFF (left panel) and $\Lambda$CDM (right panel) models. Each panel shows three curves corresponding to the different combinations of observational datasets used in the fitting: LT, LT+BAO+BBN, and LT+BAO+CMB. 
As shown in the upper panels, both models for all three dataset combinations considered follow a similar trend and remain consistent with the data over the full redshift range.
The lower panel shows the relative difference defined as $\Delta H(z) = (H_{\mathrm{EFF}}(z) - H_{\Lambda\mathrm{CDM}}(z)) / H_{\Lambda\mathrm{CDM}}(z) \times 100\%$
, which measures the deviation of the EFF model from the standard $\Lambda$CDM prediction. At low redshifts ($z \lesssim 0.5$), the relative difference $\Delta H(z)$ is negative for all three dataset combinations, indicating that the expansion rate predicted by $\Lambda$CDM is slightly larger than that of the EFF model in this regime. The curves then cross zero at $z \sim 0.5$, after which the difference becomes positive, implying a slightly faster expansion in the EFF model. The deviation increases gradually with redshift but remains below the 1\% level over the entire redshift range considered.\\

\begin{figure}[H]
	\centering
	\includegraphics[width=\columnwidth]{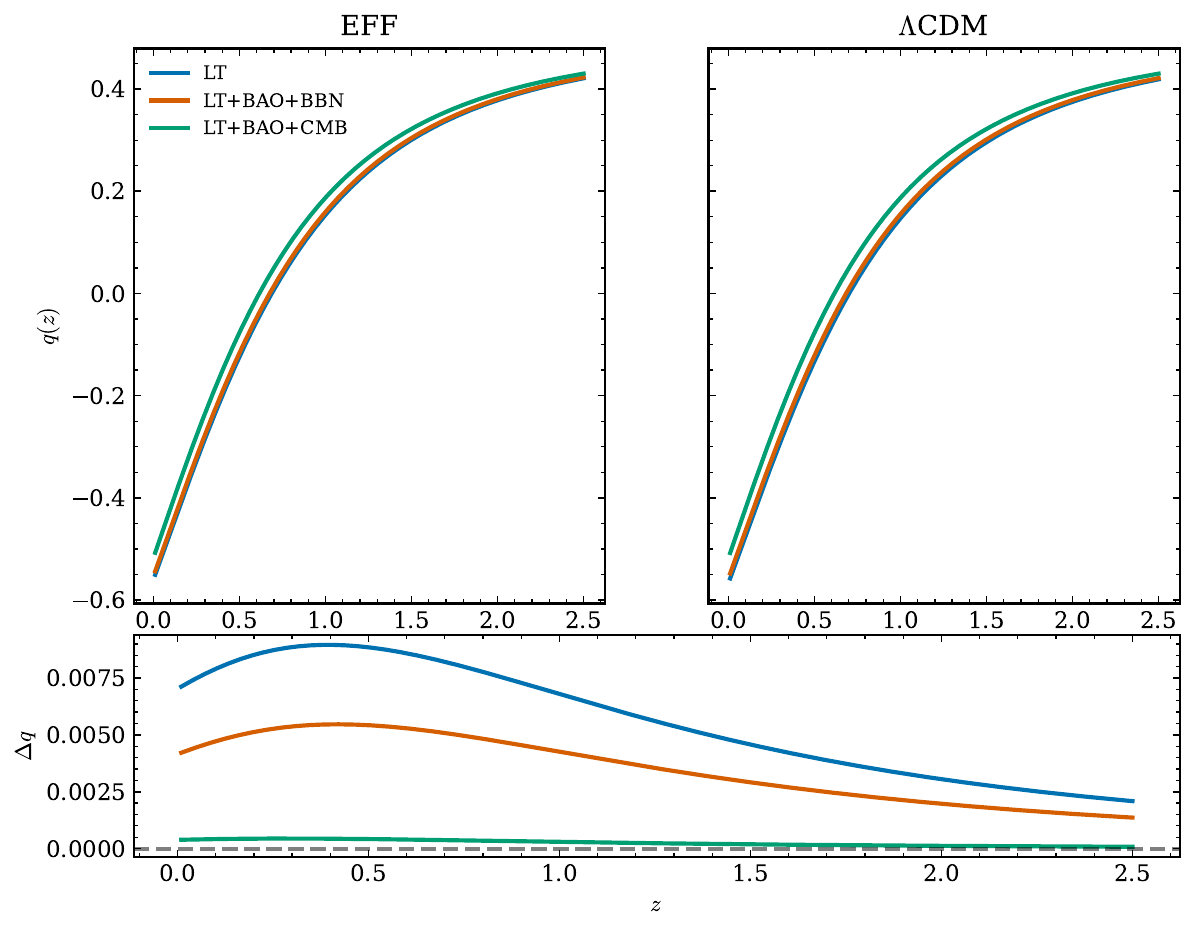}
	\caption{ The deceleration parameter $q(z)$ as a function of redshift for the $\Lambda$CDM (right panel) and EFF (left panel) cosmological models for the three dataset combinations. The lower panel shows the relative difference.}
	\label{fig:deceleration}
\end{figure}
The redshift evolution of the deceleration parameter $q(z)$ is shown in Figure ~\ref{fig:deceleration} for the $\Lambda$CDM (right panel) and EFF (left panel) models. The two models show very similar behavior over the range $0 \le z \le 2.5$.
At the present epoch, the models give
$q_0 = -0.55$
for $\Lambda$CDM and
$q_0 = -0.54$
for the EFF model, indicating a slightly weaker acceleration in the EFF model. The transition redshift, defined by $q(z_{\mathrm{tr}})=0$, is nearly the same in both cases, with
$z_{\mathrm{tr}} = 0.65.$
The lower panel shows the difference
$\Delta q(z)= q_{\mathrm{EFF}}(z)-q_{\Lambda\mathrm{CDM}}(z)$,
which remains small over the whole redshift range. The maximum deviation occurs around $z\sim0.3$, where
$\Delta q(z)\approx0.007$
for the LT dataset. The EFF model closely follows the $\Lambda$CDM prediction, with only a small positive shift in $q(z)$ at low redshift.\\

\begin{figure}[H]
	\centering
	\includegraphics[width=\columnwidth]{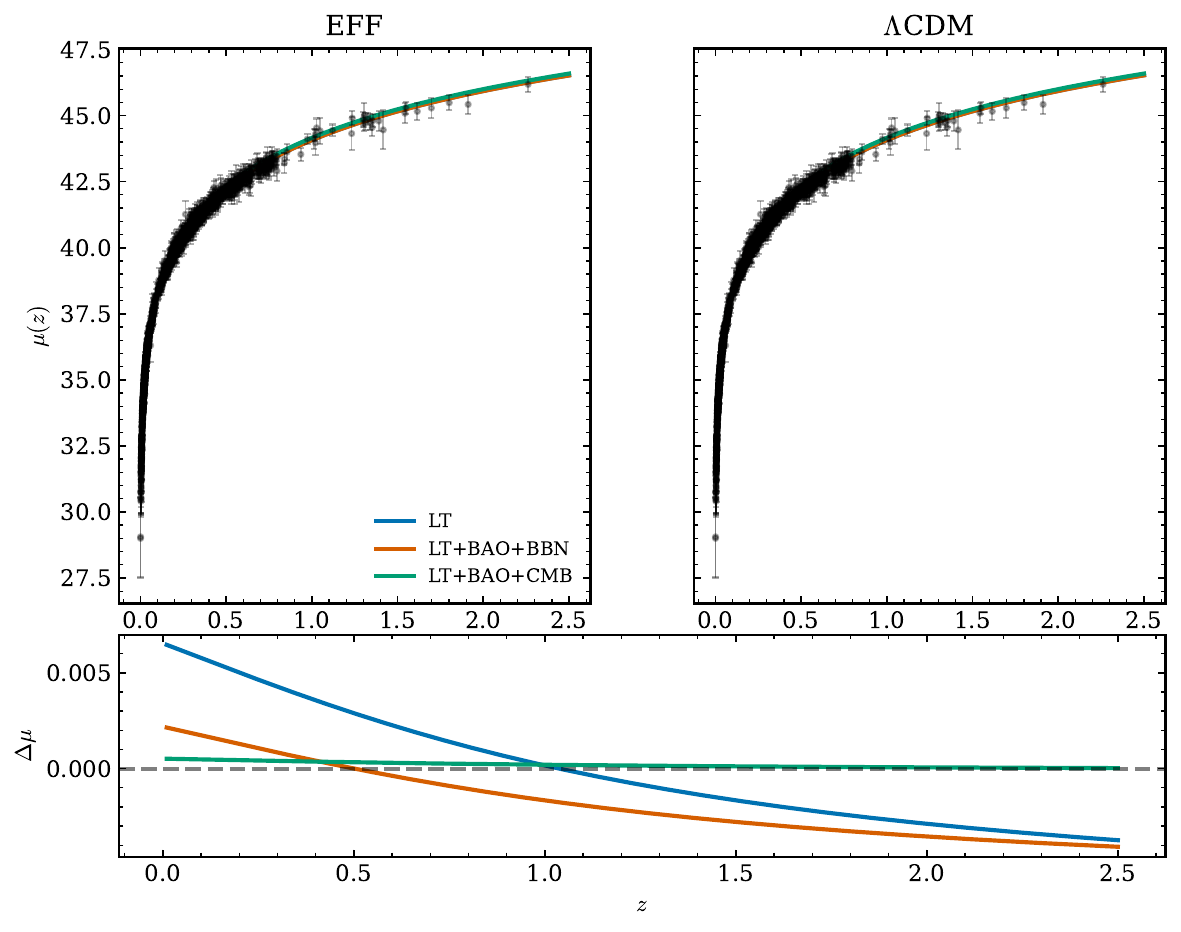}
	\caption{Distance modulus $\mu(z)$ as a function of redshift for $\Lambda$CDM (right panel) and EFF (left panel), compared to PantheonPlus data points for the three dataset combinations. The lower panel shows the relative difference.}
	\label{fig:mu}
\end{figure}
Figure~\ref{fig:mu} shows the distance modulus $\mu(z)$ for the EFF model (left panel) and the $\Lambda$CDM model (right panel). For each model, we display the predictions obtained from the LT, LT+BAO+BBN, and LT+BAO+CMB datasets. The upper panels demonstrate that all three parameter combinations provide a good fit to the supernova observations over the full redshift interval $0 \le z \le 2.5$.
The lower panel shows the difference $\Delta\mu(z) = \mu_{\mathrm{EFF}} - \mu_{\Lambda\mathrm{CDM}}$. For the dataset including CMB, the difference between the two models is nearly zero across the entire redshift range. For the other two combinations, the deviation starts with a small positive value at $z \approx 0$ and decreases as redshift increases, crossing the zero axis (at $z \approx 0.5$ for LT+BAO+BBN and $z \approx 1$ for LT) and then becoming slightly negative at higher redshifts. Despite these shifts, the absolute magnitude of the difference remains extremely small ($|\Delta\mu| < 0.007$), confirming the high degree of agreement between EFF and the standard model in describing the luminosity distance.

\begin{figure}[H]
	\centering
	\includegraphics[width=\columnwidth]{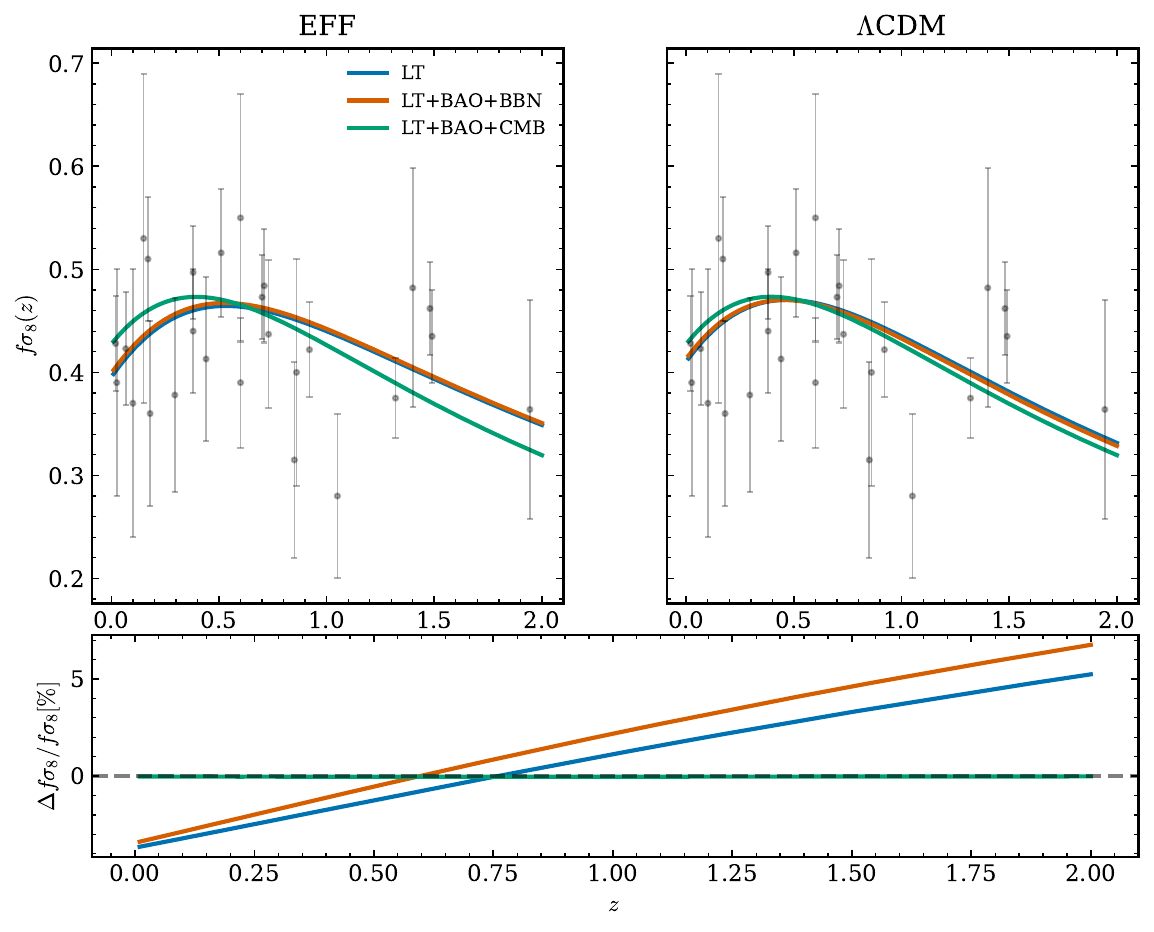}
	\caption{Evolution of $f\sigma_8(z)$ as a function of redshift for the EFF model (left panel) and the $\Lambda$CDM model (right panel), compared with observational measurements (black points). The lower panel shows the relative difference.
	}
	\label{fig:fs8}
\end{figure}
Figure~\ref{fig:fs8} displays the evolution of the growth rate $f\sigma_8(z)$ for the EFF (left panel) and $\Lambda$CDM (right panel) models, compared with observational $f\sigma_8$ data points shown as gray dots with error bars. The curves correspond to the three dataset combinations LT (blue), LT+BAO+BBN (orange), and LT+BAO+CMB (green).
The lower panel shows the relative difference 

\begin{equation}\nonumber
	\Delta(f\sigma_8) = ((f\sigma_8)_{\mathrm{EFF}} - (f\sigma_8)_{\Lambda\mathrm{CDM}}) / (f\sigma_8)_{\Lambda\mathrm{CDM}} \times 100\%.
\end{equation}

For the combination including CMB data (green line), the two models are identical, with the relative difference near zero across the entire redshift range. For the LT and LT+BAO+BBN cases, the EFF model predicts a slightly lower growth rate at low redshifts, starting at approximately $-4\%$ and $-3.5\%$ at $z=0$, respectively. These deviations diminish as redshift increases, crossing the zero-axis at $z \approx 0.6$ for LT+BAO+BBN and $z \approx 0.8$ for LT. At higher redshifts, the EFF model predicts a higher growth rate compared to $\Lambda$CDM.

\section{Conclusions}\label{con}

In this work, we place observational constraints on the Emergent Fractional Fractal (EFF) cosmological model using three datasets: LT (PantheonPlus + $H(z)$ + $f\sigma_8$), LT+DESI DR2 BAO+BBN, and LT+DESI DR2 BAO+CMB. Our main findings can be summarized as follows.
Current observations place very tight constraints on the fractal dimension parameter. With the inclusion of CMB distance priors, deviations from the $\Lambda$CDM limit are constrained to the level of $\mathcal{O}(10^{-4})$.
LT probes alone allow for mild parameter degeneracies. For the LT dataset, the parameter $d$ shows correlations with $H_0$ and $\sigma_8$, reflecting the modified matter and radiation scalings in the EFF framework. The addition of BAO+BBN data partially reduces these degeneracies, while the inclusion of CMB distance priors significantly tightens the constraints and sharply reduces the allowed parameter space.
The growth of cosmic structure provides an important independent constraint. Since the modified continuity equation directly affects the growth rate of matter perturbations, measurements of $f\sigma_8$ play a key role in constraining possible deviations in $d$, while also providing sensitivity to the amplitude of matter fluctuations $\sigma_8$.
Model comparison does not indicate any statistical preference for departures from the $\Lambda$CDM model. The AIC values indicate comparable performance between the EFF and $\Lambda$CDM models, while the BIC favors $\Lambda$CDM.

Overall, the EFF model remains consistent with current cosmological observations. However, the data strongly constrain any fractional deviation from standard cosmological dynamics and effectively drive the model very close to the $\Lambda$CDM limit.
Finally, while we have treated $d$ as a constant in this study, future work could explore the possibility of a redshift-dependent fractal dimension, which might provide a more flexible description of the cosmological evolution across different epochs.

\section{Acknowledgments}
R.J. acknowledges financial support from the Iran National Science Foundation, Iran-INSF, Grant no. 4042695.

\vspace{.3cm}
\section*{References}

\begin{thebibliography}{10}
\providecommand{\eprint}[2][]{\url{#2}}

\bibitem{Reuter:1998}
Reuter M 1998 {\em Phys. Rev. D\/} {\bf 57} 971--985 (\textit{Preprint} \eprint{hep-th/9605030})

\bibitem{Rovelli:2008}
Rovelli C 2008 {\em Living Rev. Rel.\/} {\bf 11} 5

\bibitem{Zwiebach:2004}
Zwiebach B 2006 {\em {A first course in string theory}\/} (Cambridge University Press) ISBN 978-0-521-83143-7, 978-0-511-20757-0

\bibitem{Jalalzadeh:2025}
Jalalzadeh R, Jalalzadeh S and Heydarzade Y 2025 {\em Nucl. Phys. B\/} {\bf 1017} 116945 (\textit{Preprint} \eprint{2504.04502})

\bibitem{Calcagni:2010}
Calcagni G 2010 {\em JHEP\/} {\bf 03} 120 (\textit{Preprint} \eprint{1001.0571})

\bibitem{Hawking:1976}
Hawking S~W 1976 {\em Phys. Rev. D\/} {\bf 14} 2460--2473

\bibitem{Moniz:2020emn}
Moniz P~V and Jalalzadeh S 2020 {\em Mathematics\/} {\bf 8} 313 (\textit{Preprint} \eprint{2003.01070})

\bibitem{Jalalzadeh:2022}
Jalalzadeh S 2022 {\em Phys. Lett. B\/} {\bf 829} 137058 (\textit{Preprint} \eprint{2203.09968})

\bibitem{Chen:2024}
Chen D~M and Wang L 2024 {\em Universe\/} {\bf 10} 333 (\textit{Preprint} \eprint{2409.02954})

\bibitem{JalalzadehMoniz:2022}
Jalalzadeh S and Vargas~Moniz P 2022 {\em {Challenging Routes in Quantum Cosmology}\/} (World Scientific) ISBN 978-981-4415-06-4

\bibitem{JalalzadehCosta:2022}
Jalalzadeh S, Costa E~W~O and Moniz P~V 2022 {\em Phys. Rev. D\/} {\bf 105} L121901 (\textit{Preprint} \eprint{2206.07818})

\bibitem{Calcagni:2017}
Calcagni G 2017 {\em Phys. Rev. D\/} {\bf 96} 046001 (\textit{Preprint} \eprint{1705.01619})

\bibitem{Costa:2023}
Costa E~W~d~O, Jalalzadeh R, da~Silva~J{\'u}nior P~F, Rasouli S~M~M and Jalalzadeh S 2023 {\em Fractal and Fractional\/} {\bf 7} 854

\bibitem{Jalalzadeh:2024}
Jalalzadeh R, Jalalzadeh S, Jahromi A~S and Moradpour H 2024 {\em Phys. Dark Univ.\/} {\bf 44} 101498 (\textit{Preprint} \eprint{2404.06986})

\bibitem{ElNabulsi:2013}
El-Nabulsi A~R 2013 {\em Indian J. Phys.\/} {\bf 87} 835--840

\bibitem{Rasouli:2022}
Rasouli S~M~M, Costa E~W~O, Moniz P~V and Jalalzadeh S 2022 {\em Fractal Fract.\/} {\bf 6} 655 (\textit{Preprint} \eprint{2210.00909})

\bibitem{Benetti:2024}
Benetti F, Lapi A, Gandolfi G and Liberati S 2024 {\em Class. Quant. Grav.\/} {\bf 41} 175010 (\textit{Preprint} \eprint{2407.16787})

\bibitem{Landim:2021}
Landim R~G 2021 {\em Phys. Rev. D\/} {\bf 103} 083511 (\textit{Preprint} \eprint{2101.05072})

\bibitem{Giusti:2020}
Giusti A 2020 {\em Phys. Rev. D\/} {\bf 101} 124029 (\textit{Preprint} \eprint{2002.07133})

\bibitem{daSilvaJunior:2023}
da~Silva~J{\'u}nior P, de~Oliveira~Costa E and Jalalzadeh S 2023 {\em Eur. Phys. J. Plus\/} {\bf 138} 1--16 (\textit{Preprint} \eprint{2309.12478})

\bibitem{Padmanabhan:2012}
Padmanabhan T 2012  (\textit{Preprint} \eprint{1206.4916})

\bibitem{Cai:2012}
Cai R~G 2012 {\em JHEP\/} {\bf 11} 016 (\textit{Preprint} \eprint{1207.0622})

\bibitem{Tu:2013}
Tu F~Q and Chen Y~X 2013 {\em JCAP\/} {\bf 05} 024 (\textit{Preprint} \eprint{1303.5813})

\bibitem{Hashemi:2015}
Hashemi M, Jalalzadeh S and Vasheghani~Farahani S 2015 {\em Gen. Rel. Grav.\/} {\bf 47} 53 (\textit{Preprint} \eprint{1308.2383})

\bibitem{Yuan:2013}
Yuan F~F and Huang Y~C 2013  (\textit{Preprint} \eprint{1304.7949})

\bibitem{Moradpour:2016}
Moradpour H 2016 {\em Int. J. Theor. Phys.\/} {\bf 55} 4176--4184 (\textit{Preprint} \eprint{1601.05014})

\bibitem{Chen:2022}
Chen G~R 2022 {\em Eur. Phys. J. C\/} {\bf 82} 532

\bibitem{DESI:2025}
Abdul~Karim M, Aguilar J, Ahlen S, Alam S, Allen L, Prieto C~A, Alves O, Anand A, Andrade U, Armengaud E {\em et~al.\/} 2025 {\em Phys. Rev. D\/} {\bf 112} 083515

\bibitem{Scolnic:2022}
Scolnic D {\em et~al.\/} 2022 {\em Astrophys. J.\/} {\bf 938} 113 (\textit{Preprint} \eprint{2112.03863})

\bibitem{Tamri:2026}
Tamri Z, Aghamohammadi A, Golanbari T and Khodam-Mohammadi A 2026 {\em Eur. Phys. J. C\/} {\bf 86} 96

\bibitem{Mohebi:2026}
Mohebi R, Saaidi K, Golanbari T and Karami K 2026 {\em JHEAp\/} {\bf 53} 100648 (\textit{Preprint} \eprint{2508.20129})

\bibitem{chen:2019}
Chen L, Huang Q~G and Wang K 2019 {\em JCAP\/} {\bf 02} 028 (\textit{Preprint} \eprint{1808.05724})

\bibitem{pitrou2018precision}
Pitrou C, Coc A, Uzan J~P and Vangioni E 2018 {\em Phys. Rept\/} {\bf 754} 1--66

\bibitem{Riesz:1949}
Riesz M 1949 {\em Acta Mathematica\/} {\bf 81} 1 -- 222

\bibitem{Tarasov:2018}
Tarasov V~E 2018 {\em Adv. High Energy Phys.\/} {\bf 2018} 7612490 (\textit{Preprint} \eprint{1805.08566})

\bibitem{laskin2002fractional}
Laskin N 2002 {\em Phys. Rev. E\/} {\bf 66} 056108 (\textit{Preprint} \eprint{quant-ph/0206098})

\bibitem{foreman2013emcee}
Foreman-Mackey D, Hogg D~W, Lang D and Goodman J 2013 {\em Publ. Astron. Soc. Pac.\/} {\bf 125} 306--312

\bibitem{Gelman:1992zz}
Gelman A and Rubin D~B 1992 {\em Statist. Sci.\/} {\bf 7} 457--472

\bibitem{Lewis:2019}
Lewis A 2019 {\em arXiv e-prints\/} (\textit{Preprint} \eprint{1910.13970})

\bibitem{aver2022comprehensive}
Aver E, Berg D~A, Hirschauer A~S, Olive K~A, Pogge R~W, Rogers N~S, Salzer J~J and Skillman E~D 2022 {\em Mon. Not. Roy. Astron. Soc.\/} {\bf 510} 373--382

\bibitem{Mehrabi:2015kta}
Mehrabi A, Basilakos S, Malekjani M and Davari Z 2015 {\em Phys. Rev. D\/} {\bf 92} 123513 (\textit{Preprint} \eprint{1510.03996})

\bibitem{Batista:2013oca}
Batista R~C and Pace F 2013 {\em JCAP\/} {\bf 06} 044 (\textit{Preprint} \eprint{1303.0414})

\end{thebibliography}


\end{document}